# Electron-infrared phonon coupling in ABC trilayer graphene

Xiaozhou Zan[1,2,†], Xiangdong Guo[3,4,†], Aolin Deng[6], Zhiheng Huang[1,2], Le Liu[1,2], Fanfan Wu[1,2], Yalong Yuan[1,2], Jiaojiao Zhao[1,2], Yalin Peng[1,2], Lu Li[1,2], Yangkun Zhang[1,2], Xiuzhen Li[1,2], Jundong Zhu[1,2], Jingwei Dong[1,2], Dongxia Shi[1,2,5], Wei Yang[1,2,5], Xiaoxia Yang[3,4], Zhiwen Shi[6], Luojun Du[1,2*], Qing Dai[3,4*] & Guangyu Zhang[1,2,5*]

[1] *Beijing National Laboratory for Condensed Matter Physics and Institute of Physics, Chinese Academy of Sciences, Beijing, China.*

[2] *School of Physical Sciences, University of Chinese Academy of Sciences, Beijing, China.*

[3] *CAS Key Laboratory of Nanophotonic Materials and Devices, CAS Key Laboratory of Standardization and Measurement for Nanotechnology, CAS Center for Excellence in Nanoscience, National Center for Nanoscience and Technology, Beijing, China.*

[4] *Center of Materials Science and Optoelectronics Engineering, University of Chinese Academy of Sciences, Beijing, China.*

[5] *Songshan Lake Materials Laboratory, Dongguan, Guangdong, China.*

[6] *Key Laboratory of Artificial Structures and Quantum Control (Ministry of Education), Shenyang National Laboratory for Materials Science, School of Physics and Astronomy, Shanghai Jiao Tong University, Shanghai, China.*

[†] *The authors contributed equally to this work.*

*Corresponding authors: luojun.du@iphy.ac.cn; daiq@nanoctr.cn; gyzhang@iphy.ac.cn

**Stacking order plays a crucial role in determining the crystal symmetry and has significant impacts on electronic, optical, magnetic, and topological properties. Electron-phonon coupling, which is central to a wide range of intriguing quantum phenomena, is expected to be intricately connected with stacking order. Understanding the stacking order-dependent electron-phonon coupling is essential for understanding peculiar physical phenomena associated with electron-phonon coupling, such as superconductivity and charge density waves. In this study, we investigate the effect of stacking order on electron-infrared phonon coupling in graphene trilayers. By using gate-tunable Raman spectroscopy and excitation frequency-dependent near-field infrared nanoscopy, we show that rhombohedral ABC-stacked trilayer graphene has a significantly stronger electron-infrared phonon coupling strength than the Bernal ABA-stacked trilayer graphene. Our findings provide novel insights into the superconductivity and other fundamental physical properties of rhombohedral ABC-stacked trilayer graphene, and can enable nondestructive and high-throughput imaging of trilayer graphene stacking order using Raman scattering.**

Stacking order is a unique structural degree of freedom presented in two-dimensional (2D) layered materials. It plays a significant role in governing the symmetry breaking and a wide range of fascinating electronic, optical, magnetic, and topological phenomena. For instance, 3R-stacked transition metal dichalcogenides (TMDs) lack spatial inversion symmetry and exhibit intriguing valley physics and second-order nonlinear responses such as electric-dipole allowed second harmonic generation. In contrast, inversion symmetry is restored in 2H-stacked TMD bilayers, resulting in nil valley and nonlinear responses [1-2]. ABA-stacked (Bernal) and ABC-stacked (rhombohedral) trilayer graphene have distinct electronic and optical properties [3-10]. ABC-stacked trilayer graphene exhibits an electric-field-tunable band gap and features van Hove singularities at or near the band edge, where the density of states diverges. It can also exhibit tunable Mott insulator [11], superconductor [12-13], and ferromagnetic [14-15] behavior, while these phenomena have not been observed in ABA-stacked trilayer graphene [16-17].

Electron-phonon coupling is a fundamental interaction between elementary excitations that plays a significant role in a variety of physical phenomena and quantum phase transitions in condensed matter physics. For example, electron-phonon coupling sets the intrinsic limit of electron mobility and leads to ultra-low thermal conductivity [18-21]. It also lays a foundation for charge density waves [22-23], electron hydrodynamics [24-26], superfluidity [27-28] and superconductivity [29-31]. As both the electronic band structures and phonon dispersions depend strongly on stacking geometry, a stacking order-governed electron-phonon coupling is expected in principle. Understanding the dependence of electron-phonon coupling on stacking order is fundamentally important for both comprehending a rich variety of peculiar physical phenomena and engineering novel device applications.

In this study, we report the observation of stacking order-governed electron-infrared phonon coupling in trilayer graphene. Specifically, we demonstrate that ABC-stacked trilayer graphene has much stronger electron-infrared phonon coupling strength than the ABA-stacked trilayer graphene. Our findings are supported by gate-tunable Raman spectroscopy and excitation frequency-dependent near-field infrared spectroscopy. The observed giant electron-infrared phonon coupling in ABC-stacked trilayer graphene sheds new light on its peculiar quantum properties, such as strong correlation, superconductivity, and ferromagnetism. Moreover, the distinct electron-infrared phonon coupling between ABC-stacked and ABA-stacked trilayer graphene offers a nondestructive and high-throughput imaging approach based on Raman scattering for identifying the stacking order of trilayer graphene.

**The stacking order of trilayer graphene**

Trilayer graphene possesses two distinct stacking geometries that have different symmetries and electronic properties. The ABC trilayer graphene is centrosymmetric and semiconducting, while the ABA trilayer graphene is non-centrosymmetric and semimetallic (Figs 1a and 1b). This unique characteristic of trilayer graphene makes it an exciting platform for studying the stacking order-driven electron-phonon coupling. Both ABC and ABA trilayer graphene can be directly obtained by the mechanical exfoliation from bulk crystals as they belong to structurally metastable/stable carbon allotropes. A representative trilayer graphene region with an area greater than 5000 $\mu m^2$ (highlighted by the dashed black line) on 285nm-SiO$_2$/Si substrates can be seen in the white-light microscopic image (Fig. 1c). However, having the same optical contrast, the ABC and ABA trilayer graphene can not be distinguished from the optical image. Scanning near-field optical microscope (SNOM), on the other hand, can sensitively probe the electronic band structure and optical conductivity of samples, offering an effective technique to resolve the stacking order of graphene [32-35]. Fig. 1d shows a typical SNOM image taken from a trilayer graphene with excitation frequency of 940 cm$^{-1}$. From it, the ABC and ABA trilayer graphene can be clearly distinguished as dark and bright regions, respectively.

**Stronger electron-infrared phonon coupling in the ABC trilayer graphene**

Due to the interlayer coupling and stacking geometry, high-frequency optical phonon modes in graphene couple with different valence and conduction bands, resulting in different electron-phonon interaction strengths. The ABA and ABC trilayer graphene have different point group symmetries $D_{3h}$ and $D_{3d}$, respectively, corresponding to different mode symmetries: 2E' +E" for ABA trilayer graphene, and $2E_g +E_u$ for the ABC trilayer graphene, as shown in Fig. 2a. The E' mode is both Raman and infrared active, the E" and $E_g$ modes are Raman active, $E_u$ is infrared active, and the antisymmetric vibrational mode $E_u$ can be simultaneously Raman active through inversion symmetry breaking [36-39]. Fig. 2b shows the typical Raman spectra of ABC (red line) and ABA trilayer graphene (blue line), excited by 532 nm (2.33 eV) laser at a temperature of 10 K. The ABA trilayer graphene shows a single Raman peak at ∼1588 cm$^{-1}$, assigned as Raman phonon G mode. In striking contrast, ABC trilayer graphene clearly shows an additional low wave number peak at ∼1572 cm$^{-1}$, which is absent in ABA trilayer graphene. This characteristic Raman peak is also confirmed by measurements under 633nm-laser excitation and at 300 K (refer to Fig. S1). The new Raman peak at ∼1572 cm$^{-1}$ may stem from an infrared active phonon mode. To confirm this hypothesis, we carried out far-field infrared spectroscopy measurements and the corresponding spectra are shown in Fig. 2c. Remarkably, the infrared absorption peak matches the low wave number Raman peak, suggesting that this low wavenumber Raman peak is the infrared active phonon mode.

As mentioned above, this infrared active phonon mode requires the breaking of the inversion symmetry of ABC trilayer graphene. To confirm this, we further performed gate-tunable Raman spectroscopy. Fig. 2d displays the Raman spectra of ABC trilayer graphene at gate voltage of 60 V (top panel), 0 V (middle panel) and –60 V (low panel). Besides, the Raman spectra of ABA trilayer graphene and ABC trilayer graphene at other gate voltages are also shown in Fig. S2. Notably, the frequencies of phonon G mode of both ABA and ABC trilayer graphene are strongly gate voltages dependent. Fig. 2e presents the phonon energy extracted by Lorentzian fitting as a function of gate voltages. Note that the Dirac point of our trilayer graphene device is at a gate voltage of >80 V (See Figure S3), the hole density is gradually increased as gate voltages sweep from 80 V to –60 V. With increasing the hole density, the mode of ABA trilayer graphene and high wavenumber mode of ABC trilayer graphene harden, indicating both are symmetric Raman G mode. In marked contrast, the low wavenumber component of ABC trilayer graphene softens with doping density, confirming the antisymmetric infrared active nature.

The observation of infrared active phonon mode in centrosymmetric ABC trilayer graphene by Raman spectroscopy is quite surprising, since the parity is a conserved quantity, and thus Raman and infrared transitions are mutually exclusive. One possible reason is that the inversion symmetry of ABC trilayer graphene is broken by dielectric environment doping, enabling the infrared active phonon mode Raman active. Indeed, our transport measurements show that the Dirac point is far shifted to a gate voltage of > 80V, indicating strong hole doping effect of dielectric environments. Meanwhile, in the light of third-order time-dependent perturbation theory, the intensity of first-order Raman process is given by [40-41]:

$$I(E_{laser}, \omega) \propto |M_{elop}(j,b)M_{elph}(b,a)M_{elop}(a,j)|^2$$

where $M_{elop}$ is an electron-photon matrix element between the ground and excited states and $M_{elph}$ is an electron-phonon matrix element from the initial state to the final state coupled by the G-band phonon. Since the Raman intensity is proportional to the electron-phonon coupling strength, the observation of infrared active phonon mode indicates that the electron-infrared phonon coupling is stronger in ABC trilayer graphene.

**Resonance with the infrared active phonon in ABC trilayer graphene**

We performed SNOM measurements using lasers of different frequencies to confirm the strong electron-infrared phonon coupling in ABC trilayer graphene. As shown in Figs. 3a and 3c, when the laser frequency is not at the phonon resonance frequency, the SNOM experiment reveals a low scattering amplitude intensity in ABC trilayer graphene compared to ABA trilayer graphene, indicating weak conductivity. However, when the excitation laser frequency is resonant with the infrared active phonon, the near-field signal should be dominated by the infrared phonon response if the electron-infrared phonon coupling is strong. Fig. 3b shows the SNOM image excited by laser radiation resonant with the infrared active phonon at 1573cm$^{-1}$. Electron-phonon interactions primarily arise from the coupling between phonon resonances and interband transitions. Strong interband resonances can transfer oscillator strength to phonon resonances, leading to significant enhancement of the latter. These results suggest that the electron-infrared phonon coupling strength in ABC trilayer graphene is significantly stronger than that in ABA trilayer graphene. Notably, recent studies have revealed superconductivity in ABC trilayer graphene [12-13] but not in ABA trilayer graphene [16-17]. The strong electron-infrared phonon coupling in ABC trilayer graphene may provide new insights into its superconductivity, as electron-phonon coupling may play a significant role.

**Identifying the stacking order based on the electron-infrared phonon coupling**

Previously, techniques for identifying the stacking order of graphene include SNOM [32-35], second harmonic Generation (SHG) [42], scanning Kelvin probe microscopy (SKPM) [43] and conductive atomic force microscopy (CFM) [44-45]. Thanks to the activated infrared phonon in ABC trilayer graphene, Raman spectroscopy provides an alternative and high-throughput technique for imaging the stacking order of trilayer graphene. Fig. 4a shows an optical image of mechanically exfoliated trilayer graphene. Fig. 4b presents the SNOM image taken from the trilayer graphene within the dotted square area shown in Fig. 4a. ABC and ABA trilayer graphene can be clearly identified by the dark and bright regions, respectively. Fig. 4c shows the Raman intensity mapping of infrared active phonon mode taken from the same region as in Fig. 4b. Obviously, the Raman mapping matches well with the SNOM mapping in terms of distinguishing the ABC and ABA regions. The Raman mapping provides a fast, simple, non-destructive, and high-throughput technology for identifying the stacking order of multilayer graphene.

In summary, our research has shown that the electron-infrared phonon coupling strength in ABC trilayer graphene is significantly stronger than in ABA trilayer graphene. This discovery provides new perspectives on the understanding of superconductivity and other physical properties of ABC trilayer graphene. Additionally, the differences in electron-infrared phonon coupling between ABC and ABA graphene can be utilized to develop a novel, non-destructive, and high-throughput Raman scattering imaging technique for identifying the stacking order of multilayer graphene.

## Methods

### Device fabrication

The few-layer graphene flakes were mechanically exfoliated by tape from the graphite bulk crystal, and then transferred to a substrate of 285nm-$SiO_2$/Si ($P^{++}$). In order to remove adsorbents on the substrate surface and ensure cleanliness of the interface, the $SiO_2$/Si ($P^{++}$) substrate was ultrasonically cleaned for 2 hours in acetone and deionized water before the process of mechanical exfoliation. Then, the transferred graphene samples were subjected to high-temperature annealing in an Ar/$H_2$ atmosphere for 12 hours to ensure a clean surface of the graphene samples. In order to avoid surface contamination of graphene samples by photoresist and ensure high optical quality, electrode contact was carried out by coating silver glue on the edge of the sample. Directly applied gate voltage through Si ($P^{++}$) substrate and used 285nm-$SiO_2$ as the dielectric layer.

### Raman measurement

Raman spectra is acquired using a micro-Raman spectrometer (Horiba LabRAM HR Evolution) in a confocal backscattering geometry. A solid-state laser at 532nm/633nm is focused onto the samples along the z direction by a ×100 objective. The backscattered signal is collected by the same objective and dispersed by a 600-groove mm$^{-1}$ grating. The laser power during Raman measurement is kept below 100$\mu$W in order to avoid sample damage and excessive heating. The integration time is 10s. Low temperature (10K) measurements were performed using a closed-cycle optical cryostat. Electrical measurements were performed in a probe station and under dark conditions with semiconductor parameter analysers.

### Near- and far-field infrared measurement

A scattering scanning near-field optical microscope (Neaspec) equipped with a wavelength-tunable quantum cascade laser (890-2,000 cm$^{-1}$) was used to image optical near fields. A metalized cantilever atomic force microscope tip served as a scattering near-field probe. The tip was illuminated with monochromatic p-polarized infrared light from a quantum cascade laser. Far- field infrared measurements were performed by a FTIR microscopy (Thermo Fisher Nicolet iN10).

**Acknowledgements**

This work is supported by the National Key Research and Development Program of China (Grant Nos. 2021YFA1202900, 2020YFA0309600), National Science Foundation of China (NSFC, Grant Nos. 12274447, 61888102, 11834017, 1207441, 51925203, 52102160), the Strategic Priority Research Program of CAS (Grant Nos. XDB30000000) and the Key-Area Research and Development Program of Guangdong Province (Grant No. 2020B0101340001).

**Author contributions**

G.Z. supervised the project. X.Z. designed the experiments. X.Z. fabricated the devices, performed the Raman spectroscopy and transport measurements with the help of Z.H.. X.G., X.Y. and Q.D. carried out far-field infrared spectroscopy and SNOM variable frequency measurements. A.D. and Z.S. performed SNOM characterization measurements. X.Z. analyzed the data and prepared the figure. X.Z., X.G., L.D. and G.Z. wrote the paper with the input from all the authors. All other authors were involved in discussions of this work.

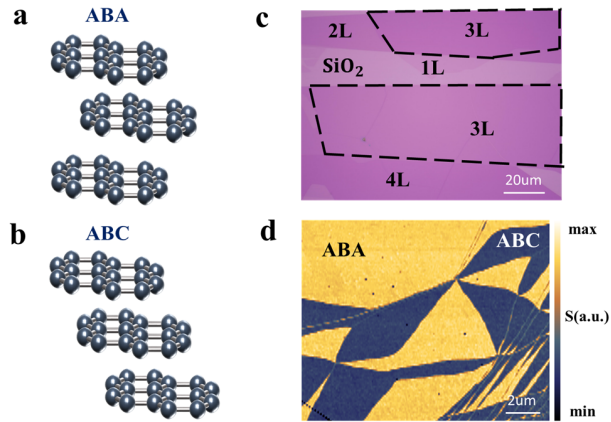

**Figure 1. The stacking order of trilayer graphene. a-b** Atomic structures of Bernal ABA-stacked and rhombohedral ABC-stacked trilayer graphene. **c** Optical microscopic image of graphene with different layers on SiO$_2$. The scale bar is 20$\mu$m. **d** Near-field infrared nanoimaging of trilayer graphene, showing Bernal ABA (bright region) and rhombohedral ABC stacking orders (dark region). The laser frequency is 940 cm$^{-1}$. The scale bar is 2$\mu$m.

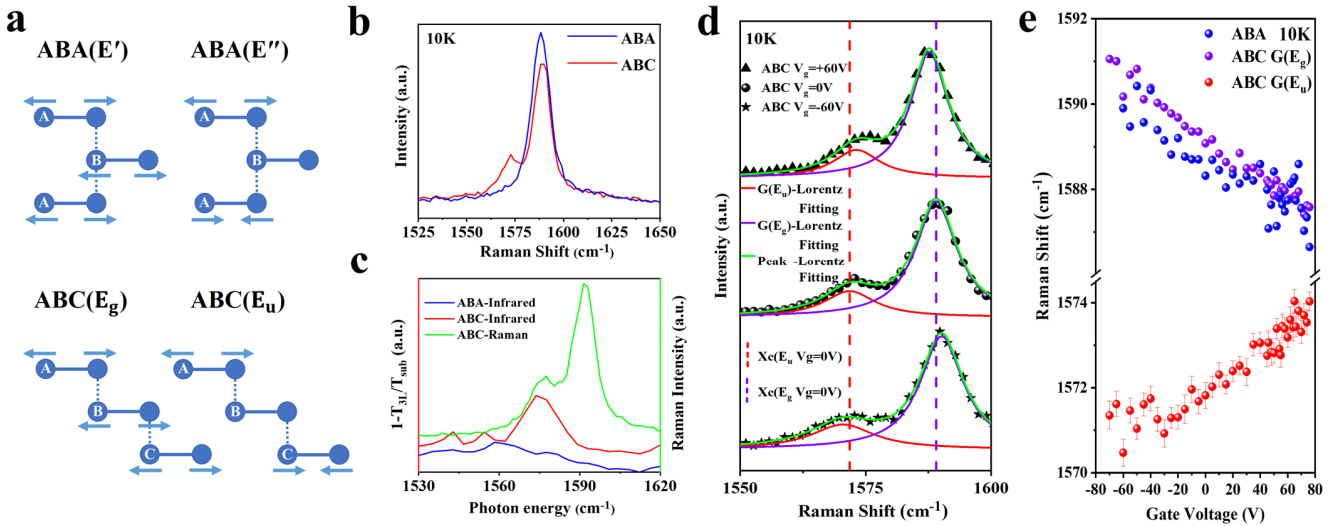

**Figure 2. Gate-tunable Raman spectra of ABA and ABC trilayer graphene. a** Schematic diagram of the typical phonon vibration modes of ABA (upper panel) and ABC trilayer graphene (lower panel). **b** Raman spectra of ABA and ABC trilayer graphene at 0 V gate voltage. **c** Far-field infrared spectra of ABA (blue line) and ABC trilayer graphene (red line) are shown on the left coordinate axis, and Raman spectrum of ABC trilayer graphene (green line) are shown on the right coordinate axis. The measurements are performed at 300K. **d** ABC trilayer graphene Raman spectra and the corresponding Lorentz fitting at 60V, 0V and -60V gate voltages. **e** The fitted phonon frequencies as a function of the applied gate voltages. The measurements are performed at 10 K.

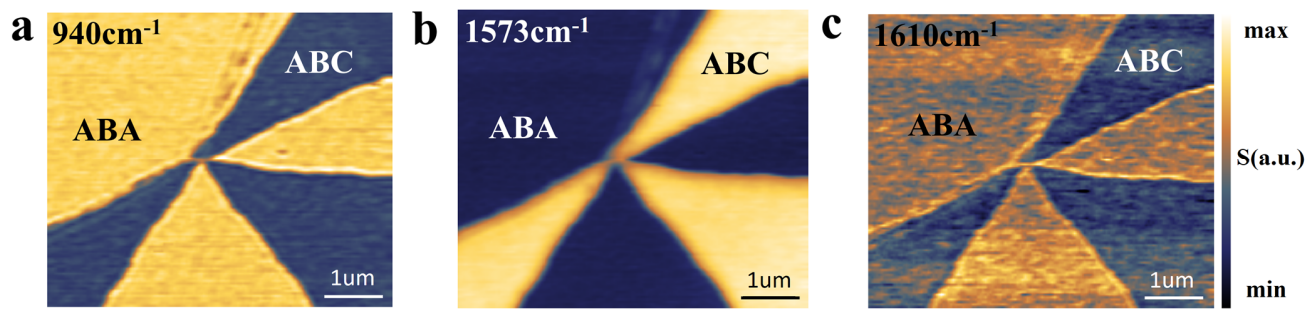

**Figure 3. Near-field infrared nano-imaging of ABA and ABC trilayer graphene at different excitation frequencies. a-c** Excitation frequencies are 940 cm$^{-1}$, 1573 cm$^{-1}$ and 1610 cm$^{-1}$, respectively. The scales are 1$\mu$m.

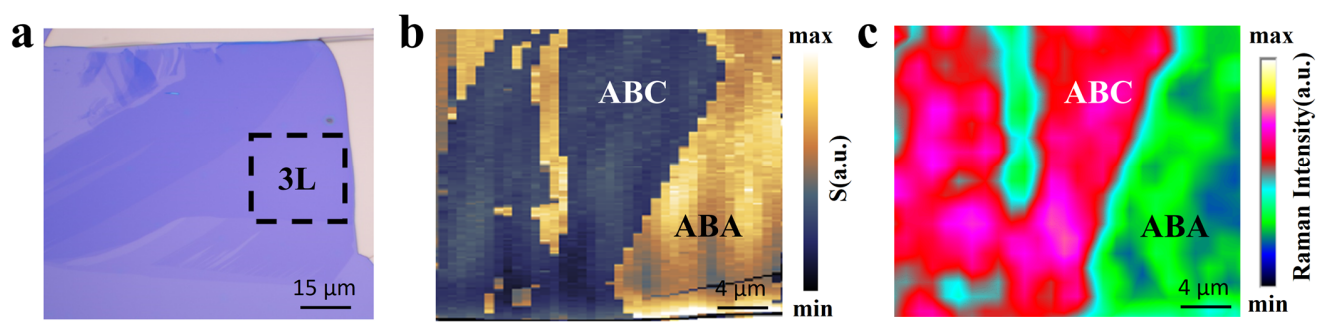

**Figure 4. Imaging the stacking order of trilayer graphene by Raman scattering. a-b** Optical microscope and SNOM diagram of trilayer graphene, the laser frequency is 940 cm$^{-1}$, where b is the area of the dotted line box in Fig. a. **c** Imaging the stacking order of trilayer graphene by Raman scattering. The mapping Raman intensity is the Raman response of ABA-stacked and ABC–stacked trilayer graphene at the corresponding wavenumber of infrared phonon. The scale is 4$\mu$m.